\documentclass[aps,prd,unsortedaddress,superscriptaddress,showpacs,nofootinbib,11pt]{revtex4-1}

\usepackage{graphicx,color}
\usepackage{relsize}
\usepackage{slashed}
\usepackage[normalem]{ulem}
\usepackage{tabu}
\usepackage{rotating}
\usepackage{bigstrut}
\usepackage{makecell}
\usepackage[dvipsnames]{xcolor}

\usepackage{amsmath}
\usepackage{amssymb,bm}
\usepackage{amsthm}
\usepackage{mathrsfs}
\usepackage{fancyhdr}
\usepackage{array}
\usepackage[all]{xy}
\usepackage{euscript}
\usepackage{enumerate}
\usepackage{slashed}
\usepackage{hyperref}
\usepackage{subfigure} 
\usepackage{mathtools}

\usepackage{soul}

\hypersetup{colorlinks=true,linkcolor=blue,citecolor=blue,menucolor=black,urlcolor=blue,filecolor=blue}

 \newcommand{\maojun}[1]{{\color{blue} #1}}

\newcommand{\swu}{\affiliation{School of Physical Science and Technology,
	Southwest University, Chongqing 400715, China}}

\begin{document}

\title{Decoding the near-threshold $X_{0,\,1}(4140)$ and $X_{1}(4685)$ states via OZI-suppressed coupled-channel scattering}

\author{Mao-Jun Yan}
\email{yanmj0789@swu.edu.cn}
\swu

\date{\today}

\begin{abstract} 
To decode the near-threshold dynamics of the $X_{0,\,1}(4140)$ and $X_{1}(4685)$ states, we investigate the OZI-suppressed $\{D_{s}\bar{D}_{s},\, J/\psi \phi,\, D_{s}^{\ast}\bar{D}_{s}^{\ast}\}$ coupled-channel scattering in $B\to D_{s}\bar{D}_{s} K$ decays using the effective range expansion. We demonstrate that the $X_{0}(4140)$, associated with a dip in the lineshape, corresponds to a dynamically generated pole near the $J/\psi \phi$ threshold. The single-channel $J/\psi \phi$ scattering length is extracted to be $1.11\pm 0.65\,\rm{fm}$, yielding an effective scattering length of $0.12^{+0.20}_{-0.10}+i0.78^{+0.20}_{-0.40} \, \rm{fm}$ when coupled channels are included. By treating the spin-spin interaction as a subleading effect, we predict a $J^{PC}=1^{++}$ virtual state near the $J/\psi \phi$ threshold, which naturally resolves the empirical ambiguities surrounding the $X_{1}(4140)$ width. Extending this framework via heavy quark spin symmetry, we further interpret the $X_{1}(4685)$ as a $\psi(2S)\phi$ hadronic molecule. Ultimately, these findings highlight how the $X(4140)$ family and $X_{1}(4685)$ serve as unique theoretical windows into the Fierz rearrangement and OZI suppression mechanisms in low-energy strong interactions.
\end{abstract}

\maketitle

\section{Introduction}
Quantum chromodynamics (QCD) is the fundamental quantum field theory describing the strong interaction, but the low energy scattering is hard to estimate from QCD perturbatively. To overcome this difficulty in dynamics, some features of the S-matrix may shed light on the low energy scattering and provide hints to reveal the QCD in the non-perturbative region, where the causality (or analyticity) is widely studied in nuclear and hadron physics. 
Therein, the effective range expansion (ERE) was initially introduced by H. Bethe \cite{Bethe:1949yr} and a simplified version is adapted to the analysis of $X(3872)$ lineshape involving coupled channel scattering \cite{Braaten:2005jj}. This analysis is developed for the near-threshold structures in the heavy spectrum, with the lineshapes determined by the near-threshold poles and the production rates \cite{Dong:2020hxe}, where the poles relate to the dynamics.

Compared to model building, the physical spectrum is more complicated and performs due to interference between Feynman diagrams, where the sharp lineshape from a single diagram may finally be smeared.  Therefore, the visible sharp lineshapes in the invariant mass spectrum are expected to be dominated by the underlying scattering dynamics, while the interference from a simple background plays an insignificant role. In the hidden charm sector, there are constraints on isospin (flavor) and spin-parity in the $D_s \bar{D}_s$ final state. These constraints filter the interference among Feynman diagrams, contributing to $D_s \bar{D}_s$ invariant mass distribution, and yield a simple background term in $D_s \bar{D}_s$ scattering. As a result, the lineshape driven by ERE is valid to study the new results on $D_s \bar{D}_s$ invariant mass distribution in B decay reported by LHCb \cite{LHCb:2022aki}.

Besides $X(3960)$ is observed in $D_s^+ D_s^-$ invariant mass distribution,  a dip around $4140\, \mathrm{MeV}$ can be described either by a $0^{++} X_0(4140)$ resonance having a significance of $3.7 \sigma$ in the baseline model, or the coupled-channel effect of the $J / \psi \phi \leftrightarrow D_s^{+} D_s^{-}$ reaction in K-matrix parameterization including a bare resonance, where 5 real parameters ($M_R$, $g^R_1$, $f_{11}$, $f_{12}$ and $f_{22}$) and 3 complex parameters ($\beta_R$, $\beta_1$ and $\beta_2$) are employed \cite{LHCb:2022aki}. 
However, a recent analysis excludes the configuration that $X_0(4140)$ is a cusp from coupled channel scattering \cite{Ding:2023yuo}. In order to understand the nature of the apparent dip, heavy quark spin symmetry is imposed to $D_s \bar{D}_s -J/\psi \phi- D_s^{\ast} \bar{D}_s^{\ast}$ scattering, where the poles in the scattering provide a hint on understanding the lineshape in $D_s\bar{D}_s$ invariant mass distribution.


Meanwhile, the $X_1(4140)$ reported by several collaborations \cite{CDF:2009jgo,LHCb:2012wyi,CMS:2013jru,D0:2015nxw,LHCb:2021uow}, is either narrow or broad in the Breit-Wigner parameterizations. On the theoretical side, this state is interpreted to be the  $D_s^{ \pm} D_s^{*\mp}$  cusp \cite{Swanson:2014tra, Karliner:2016ith,Liu:2016onn,Nakamura:2021bvs}, $s\bar{s}c\bar{s}$ state of $1^+$ in QCD sum rule \cite{Chen:2016oma}, tetraquark in diquark model \cite{Lebed:2016yvr}, and the $s\bar{s}c\bar{c}$ counterpart of
the X(3872) in the relativized diquark model \cite{Anwar:2018sol}. The nature of $X_1(4140)$ is not explicit yet. Furthermore, the $X_1(4140)$ has also been proposed as a candidate for the conventional $\chi_{c1}(3P)$ charmonium state, where estimations indicate it should be a narrow resonance. Such an assignment could be experimentally tested by searching for this state in the $B \to K \chi_{c1} \pi \pi$ decay channel \cite{Chen:2016iua}. Regarding the $X_1(4140)$ closing to $J/\psi \phi$ threshold, a better understanding of $J/\psi \phi$ scattering is meaningful for insight into $X_1(4140)$ and the $X_1(4685)$ near $\psi(2S)\phi$ threshold \cite{LHCb:2021uow}.

Before proceeding, we should notice that $D^{\ast}\bar{D}^{\ast}$, $\eta_c \eta^{\prime}$ and $\eta_c(2S)\eta$ couple to this energy region in $J^{PC}=0^{++}$ sector. Concerning $D^{\ast}\bar{D}^{\ast}$ weakly coupled to the poles near $D_s^{\left(\ast \right)}\bar{D}_s^{\left(\ast \right)}$ \cite{Ji:2022vdj}, the transition between $D^{\ast}\bar{D}^{\ast}$ and $D_s\bar{D}_s$ is attributed to be subleading order with respect to the one in elastic $D_s \bar{D}_s$ scattering counted as leading order. In the charmonium-meson scattering, the non-negligible finite charmonium widths smear the
sharp lineshape around threshold \cite{Guo:2019twa,Du:2021zzh} by rotating the pole in T-matrix to be far away from physical Riemann Sheet (RS), where $\Gamma_{\eta_c}: \Gamma_{J/\psi}=344$ and $\Gamma_{\eta_c(2S)}: \Gamma_{J/\psi}=121.5$ are evaluated with the central valued widths in Review of Particle Physics (RPP) \cite{Workman:2022ynf}
, which inspires us to exclude the coupled channel effects from $\eta_c \eta^{\prime}$ and $\eta_c(2S)\eta$ in the present work \footnote{The branch ratios of $B^+\to J/\psi \phi K^+$ and $\eta_c\eta K^+$ are $\left( 5.5 \pm 0.4 \right)\times 10^{-5}$ and $<2.2 \times 10^{-4}$, respectively. They are much smaller than the one,  $\left( 1.55 \pm 0.21 \right)\times 10^{-3}$, in $B^+ \to \bar{D}^0 D^+ K^0$. }.

This note is organized as follows: in section II, the $\{ D_s\bar{D}_s,\,J/\psi\phi,\,D_s^{\ast}\bar{D}_s^{\ast}\}$ coupled channel scattering in $J^{PC}=0^{++}$ sector is displayed in ERE with a zero range expansion that is applied to fit the observation in $D_s\bar{D}_s$ invariant mass distribution; In section III, the extracted interaction in contact range is used to analysis $\{D_s\bar{D}_s^{\ast},\,J/\psi \phi\}$ scattering in heavy quark spin symmetry, and yields predictions of lineshapes on the $X_1(4140)$; A similar analysis on $\{J/\psi \phi, \psi(2S) \phi\}$ scattering explain the newly reported $X_1(4685)$; At last, a summary is presented in Section IV.

\section{ $D_s\bar{D}_s$-$J/\psi \phi$-$D_s^{*}\bar{D}^{*}_s$ scattering in $J^{PC}=0^{++}$ sector}

\subsection{Interaction in contact range}

In low-energy scattering, the pole position of a bound state is determined by the leading order interaction, which is also adaptable for the virtual state close to the threshold. This power counting rule is followed to construct a leading order interaction around the $J/\psi \phi$ threshold. 
The leading order interaction in $D_s\bar{D}_s$ invariant mass distribution differs from previous studies on the heavy hadronic molecules, where $J/\psi \phi$ couple to the open charmed meson pairs. Therefore, it is necessary to introduce a heavy quark spin symmetry (HQSS) inspired effective field theory (EFT), which covers the OZI-allowed interaction in open charmed meson channels and the suppressed one in $J/\psi \phi$ scattering. In this framework, there are two energy scales, QCD energy scale $\Lambda_{QCD}$ and charm quark mass $m_c$, in the OZI allowed scattering, the latter of which also drives the OZI suppressed $J/\psi \phi$ scattering. 

In the classic EFT without OZI suppressed transition, $m_c$ is integrated out, and the interactions are expanded in terms of $Q/\Lambda_{QCD}$ with a soft momentum $Q$ in the scattering, where the hard scale $\Lambda_{QCD}$ is replaced by other energy scale relating to the light meson mass, e.g., pion mass $m_{\pi}$ is the hard scale in XEFT \cite{Fleming:2007rp}. The vector meson mass $m_{\rho}$ is hard in the light-meson-saturation model \cite{Peng:2020xrf,Yan:2023ttx,Qiu:2023uno, Yu:2024sqv}.
However, the inclusion of $J/\psi \phi$ gains difficulty in separating $m_c$ and $\Lambda_{QCD}$ in the coupled channel scattering. In order to get rid of the ambiguity of energy scales and present a proper power counting rule, the renormalization group (RG) evolution displays the connection in contact range interactions ($V_C$) between different scales,
\begin{eqnarray}
    \frac{d}{d \Lambda}\left\langle\Psi\left|C(\Lambda)\right| \Psi\right\rangle=0,
\end{eqnarray}
with the two-body scattering wavefunction $\Psi$.
If the wave function has a power-law behavior $\Psi(r) \sim$ $r^{\alpha / 2}$ at distances $r \sim 1 / \Lambda$, the RG equation (RGE) above leads to
\begin{eqnarray}
 \frac{C\left(\Lambda_1\right)}{\Lambda_1^\alpha}=\frac{C\left(\Lambda_2\right)}{\Lambda_2^\alpha},   
\end{eqnarray}
from which an EFT can be constructed with a single energy scale $\Lambda_0$. For the exponent $\alpha$, the semi-classical approximation is used together with the Langer correction \cite{Langer:1937qr}, leading to $\Psi(r) \sim \sqrt{r}$ or $\alpha=1$.
The RGE makes it possible to construct an EFT with a single energy scale $\Lambda_0$ in the coupled channel scattering around $J/\psi \phi$ threshold, where the hard scale $\Lambda_0=m_{\phi}$ is adapted concerning the $\phi$ meson scatters off $J/\psi$ non-relativistically.

In the elastic $D_s^{\left(\ast \right)}\bar{D}_s^{\left(\ast \right)}$ scattering with positive C-parity,  the charmed meson $D_s$ and $D_s^{\ast}$ form a spin doublet $H=\left(D_s,\, D_s^{\ast}\right)$ with charmed quark spin conserved, and the light degree of freedom $s_l=1/2$. $s_l \otimes s_l=0 \oplus 1$ indicates that two contact-ranged terms are necessary for the effective Lagrangian. 
From HQSS, we expect heavy hadron interactions to be independent of the spin of the heavy quarks within them. This symmetry is implemented at the practical level by writing down superfields that group together the ground and excited states of a heavy hadron. For the S-wave charmed mesons $D_s$ and $D^*_s$, their superfield is
\begin{eqnarray}
  H=\frac{1}{\sqrt{2}}\left[D_s+\vec{\sigma} \cdot \vec{D}^*_s\right],  
\end{eqnarray}
where the C-parities of charm mesons are $\mathcal{C}\vert D_s\rangle =\vert \bar{D}_s\rangle$ and $\mathcal{C}\vert D_s^{\ast}\rangle =\vert \bar{D}_s^{\ast}\rangle$ \cite{Hidalgo-Duque:2012rqv,Guo:2013sya,Yang:2020nrt,Dong:2021juy,Yan:2021tcp}.
It should be noted that under Heavy Quark Spin Symmetry, the $D_s$ and $D_s^*$ field operators are unified into the superfield $H$ via their light-quark degrees of freedom. In the scattering process, since we are working in the non-relativistic limit, the polarization vectors ($\vec{\epsilon}$) of the vector mesons satisfy the approximation $\epsilon_i \cdot \epsilon_j = \delta_{ij}$. Consequently, the explicit dependence on the polarizations is completely absorbed by the spin-dependent operators (the Pauli matrices) in the potential, and this approximation is consistently applied to the loop integrals in the Green's functions.

With the previous superfields, the most general S-wave contact-range Lagrangian is
\begin{eqnarray}
\mathcal{L}_{\text {c }} & =C_0 \operatorname{Tr}\left[H^{\dagger} H\right] \operatorname{Tr}\left[\bar{H}^{\dagger}\bar{H}\right] 
 +C_1 \operatorname{Tr}\left[H^{\dagger} \sigma_k H\right] \operatorname{Tr}\left[\bar{H}^{\dagger} \sigma_k \bar{H}\right],
\end{eqnarray}
where $\rm{Tr}$ traces on the flavor and $\sigma$ is Pauli matrix.
These two terms can be written as a combination of spin-independent and dependent terms in the potential,
\begin{eqnarray}  
V_{C_c}\left(q \right)&=& C_0 + C_1 \left(\vec{\sigma}_{L1} \cdot \vec{q} \right) \left(\vec{\sigma}_{L2} \cdot\vec{q}\right), \label{eq:Vc}
\end{eqnarray} 
where $C_0$ and $C_1$ are coupling constants, $\vec{\sigma}_{L 1}$ and $\vec{\sigma}_{L 2}$ are the light-spin operators for the mesons, respectively, where  $\vec{q}$ is the momentum in the center of mass frame. The momentum carried by $D_s^{\ast}$  is $\vert \Vec{q}\vert =\sqrt{m_{D_s^{\ast}}\Delta}\simeq 478\,\rm{MeV}$ with $\Delta=2m_{D_s^{\ast}}-m_{J/\psi}-m_{\phi}$, and $\frac{\vec{q}^2}{\Lambda_0^2}\simeq 0.22$, which implies that the spin-spin interaction ($C_1$) term is relatively suppressed with comparing to the spin-independent one $C_0$. Therefore, the $C_1$ is switched off in the following sections, which generates 
\begin{eqnarray}
V_{D_s\bar{D}_s,D_s\bar{D}_s}&=&V_{D_s^{\ast}\bar{D}_s^{\ast},D_s^{\ast}\bar{D}_s^{\ast}},
\end{eqnarray}
with the subscripts labeling the scattering channels. 

In the elastic $J/\psi \phi$ scattering, the effective Lagrangian contains two terms as well and reads
\begin{eqnarray}
    \mathcal{L_{\text{d}}}&=D_0 \left( J/\psi \cdot J/\psi^{\dagger}\right) \left(\phi \cdot \phi^{\dagger} \right) + D_1 \left( J/\psi \vec{S}_1 J/\psi^{\dagger}\right) \left( \phi \vec{S}_2 \phi^{\dagger}\right),
\end{eqnarray}
which yields the potential 
\begin{eqnarray}  V_{C_d}\left(p \right)&=& D_0 + D_1 \left(\vec{S}_{1} \cdot \vec{p} \right) \left(\vec{S}_{2} \cdot\vec{p}\right),
\end{eqnarray} 
with the spin-1 operators $S_{1,\,2}$ acting on vector mesons, where $\vert \vec{p}\vert $ is small, the $D_1$ term is also switched off. 

In the inelastic transitions, the contact ranged interaction, with respect to HQSS, reads
\begin{eqnarray}
    \mathcal{L}_{\text{e}}&= E_0 \bar{H}\epsilon_k J/\psi \phi \epsilon^k H, + E_1 \eta_c \eta^{(\prime)} \bar{H}H,
\end{eqnarray}
with a vanished $E_1$ corresponding to the excluded $\eta_c\eta^{(\prime)}$ channel in present study, where the transition is normalized in the spin space\maojun{.}
The spin wave function of the open charmed channels reads
\begin{eqnarray}
    \left(\begin{array}{l}
\left|D_{s} \bar{D}_{s}\right\rangle \\
\left|D_{s}^* \bar{D}_{s}^*\right\rangle
\end{array}\right)_{J=0}=\left(\begin{array}{cc}
\frac{1}{2} & \frac{\sqrt{3}}{2} \\
\frac{\sqrt{3}}{2} & -\frac{1}{2}
\end{array}\right)\left(\begin{array}{l}
|0_l \otimes 0_Q\rangle \\
|1_l \otimes 1_Q\rangle
\end{array}\right)_{J=0},
\end{eqnarray}
with subscripts $l$ and $Q$ standing for the light and heavy degrees of freedom in the scattering.
This indicates 
\begin{eqnarray}
    V_{D_s\bar{D}_s,J/\psi \phi}&= \sqrt{3}E_0,\nonumber\\
    V_{D_s^{\ast}\bar{D}_s^{\ast},J/\psi \phi}&= - E_0.
\end{eqnarray}
In addition to the $E_0$ term, there are energy-dependent contact range interactions
in the inelastic transition, which is not included and attributed to the subleading interaction. 

Meanwhile, there is another inelastic transition between charmed mesons, where the S-wave transition between $D_s\bar{D}_s$ and $D_s^{\ast}\bar{D}_s^{\ast}$ is characterized by the light degree of freedom in the scattering and defined as
\begin{eqnarray}
    V_{D_s\bar{D}_s,\,D_s^{\ast} \bar{D}_s^{\ast}}&=& \frac{\sqrt{3}}{2} \left( V_{D_s^{\ast}\bar{D}_s^{\ast}, D_s^{\ast}\bar{D}_s^{\ast}}- V_{D_s \bar{D}_s, D_s \bar{D}_s}\right).\label{eq:V13}
\end{eqnarray}
The energy-dependent transition in this process is subleading and not presented now.

These leading order interactions (V) form the EFT in the low energy $D_s \bar{D}_s$ scattering, and the nonrelativistic $T$-matrix is given by the Lippmann-Schwinger equation as
\begin{eqnarray}
T_{i j}=\left[(1-V G)^{-1} V\right]_{i j}. \label{eq:Tmatrix}
\end{eqnarray}
The interaction kernel $V$ is a matrix for constant contact terms, and $G$ is a $3 \times 3$ diagonal matrix with the diagonal matrix element $G_i$ given by the nonrelativistic two-body loop function,
\begin{eqnarray}
G_i^{\Lambda} & =\int^{\Lambda}_{0} \frac{d^3 \mathbf{q}}{(2 \pi)^3} \frac{2 \mu_i}{\mathbf{q}^2-k_i^2-i \epsilon}  =i \frac{\mu_i}{2 \pi} k_i+\frac{\mu_i}{\pi^2} \Lambda\left[1+O\left(\frac{k_i^2}{\Lambda^2}\right)\right], \label{eq: Green'sFunction}
\end{eqnarray}
with the channels' reduced masses $\mu_i$, where the numerators of vector meson propagators are not written down.
The ultraviolet (UV) divergence in the loop integral is regularized using a three-momentum cutoff $\Lambda$. In Eq. (\ref{eq:Tmatrix}), the cutoff dependence of the loop function $G_i$ is absorbed by the interaction kernel $V$, and the resulting parameters $a_{i j}$ are cutoff-independent. The coupled channel scattering amplitude around $J/\psi \phi$ threshold is expressed in terms of these parameters, 
\begin{eqnarray}
    \frac{T}{8\pi th_2}&=&\begin{pmatrix}
        \frac{1}{a_{11}}-ik_1 & \frac{1}{a_{12}} & \frac{1}{a_{13}}\\
        \frac{1}{a_{12}} & \frac{1}{a_{22}} -ik_2 & \frac{1}{a_{23}} \\
        \frac{1}{a_{13}} & \frac{1}{a_{23}} & \frac{1}{a_{33}} -ik_3
    \end{pmatrix}^{-1},\label{eq:Tmatrix2}
\end{eqnarray}
 with $i,\,j=1,\,2,\,3$ labeling channels $D_s\bar{D}_s$, $J/\psi \phi$ and $D_s^{\ast}\bar{D}_s^{\ast}$, respectively \footnote{Regarding the $D\bar{D}$ channel do not strongly couple to $X(3960)$, the imaginary part of $a_{11}$ from $D\bar{D}$ is small and counted to be subleading interaction.},
where  $a_{ij}$ are \textit{scattering length},  and $k_i$ is the c.m. momentum of the channel-$i$. It should be noted that the parameters $a_{11}$, $a_{22}$, and $a_{33}$ introduced here represent the scattering lengths for the single-channel interactions and are therefore strictly real numbers. However, after dynamically incorporating the coupled-channel effects, unitarity ensures that the effective scattering length (e.g., $a_{22}^{eff}$) naturally acquires an imaginary part corresponding to the open-channel effects.
More details on the relation between the interaction kernel and scattering length can be referred to in Appendix I.
If the interaction is strong enough, a near-threshold pole of the $T$-matrix can be generated as a zero of the determinant of the matrix in the right-hand side of Eq. (\ref{eq:Tmatrix2}).


\subsection{Production amplitude and fit}


As illustrated in Fig.~1, the production amplitude as a function of the $D_{s}\bar{D}_{s}$ invariant mass is written as
\begin{eqnarray}
     \mathcal{M}_a&=&P_1^{\Lambda} +P_1^{\Lambda} G^{\Lambda}_1 T_{11}+P_1^{\Lambda} G^{\Lambda}_2 T_{21} +P_3^{\Lambda} G^{\Lambda}_3 T_{31},
\end{eqnarray}
with the immediate $\left\{D_s \bar{D_s},\, J/\psi \phi,\, D_s^{\ast}\bar{D}_s^{\ast}  \right\}$ channels,
where $P_i^{\Lambda}$ are the bare production vertices and $G_i^{\Lambda}$ are the Green's functions in Eq. (\ref{eq: Green'sFunction}).
Concerning the difference between production rates of open-charmed and hidden-charmed channels
\cite{Du:2020bqj}, $P_{1,\,3}^{\Lambda}\gg P_{2}^{\Lambda}$ with $P_{2}^{\Lambda}=0$, leads to 
\begin{eqnarray}
    \mathcal{M}_a &=&P_1^{\Lambda} + P_1^{\Lambda} G_1^{\Lambda} T_{11} + P_3 ^{\Lambda} 
 G_3^{\Lambda} T_{31}, \nonumber\\
    &=&P_1 T_{11} + P_3 T_{31}, \label{eq:amplitude}
\end{eqnarray}
where the ($P_{1,\, 3}$) are renormalized production rates.
The cutoff dependence of the loop functions in Eq. (\ref{eq: Green'sFunction}) must be absorbed by the production vertex $P_i^{\Lambda}$. Such a requirement is fulfilled by a multiplicative renormalization with $P_{i}\propto 1 / \Lambda$ and by keeping only the leading order term (in the expansion in powers of $p_i / \Lambda$ ) in the loop function \cite{Sakai:2020psu}. To make a fit in mass region $m_{\phi}+m_{J/\psi}\pm \Delta E$ with $\Delta E=180 \rm{MeV}$, the non-relativistic $\phi$ carries a maximal three-momentum $k=\sqrt{2\mu_2 \Delta E}\sim 525 \,\rm{MeV}$ with the reduced mass $\mu_2=m_{\phi}m_{J/\psi}/\left(m_{\phi} +m_{J/\psi} \right)$, and $\delta_0=k^2/m_{\phi}^2 \sim 0.27$, which is reasonable to switch off the subleading order interaction or the effective range correction term, $r_0$ in the effective range expansion. 
Given the currently limited statistics in the $D_{s}\bar{D}_{s}$ and $J/\psi\phi$ invariant mass distributions, introducing these subleading terms—such as the spin-spin interaction ($C_{1}$) or effective range corrections ($\mathcal{O}(\delta_{0})$)—would inevitably over-parameterize the model and inflate the uncertainties of the extracted leading-order parameters. 
Therefore, to capture the dominant dynamical features, we restrict our current analysis to the leading-order interactions. 
We anticipate that the upcoming high-statistics data releases from the LHCb collaboration will allow for a stringent constraint on these higher-order effects in future studies.

\begin{figure}[htb]
	\centering
 \includegraphics[scale=0.50]{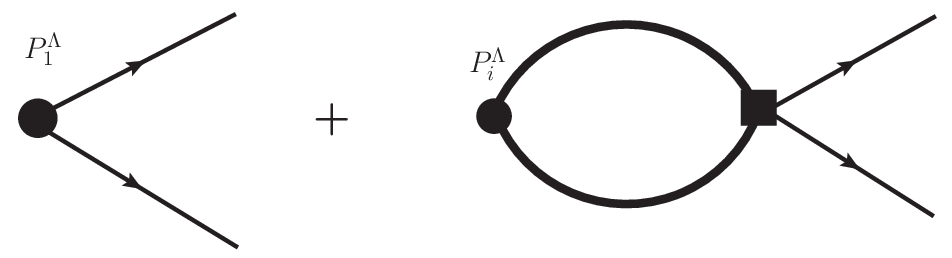}
	\caption{The Feynman diagram of the processes 
 in $D_s\bar{D}_s$ invariant mass distribution with thick and thin lines representing the intermediate and outgoing particles, respectively, in addition to the  circled  production vertex and the rectangle unitarized scattering amplitude.}\label{production}
\end{figure}

To extract the parameters in Eq. (\ref{eq:amplitude}), a fit to the $D_s \bar{D}_s$ invariant mass distribution is performed as
\begin{eqnarray}
    \frac{d N}{d m} &=& \vert \mathcal{M}_a\vert^2 \frac{\vert \vec{k}_1\vert}{8\pi m},
\end{eqnarray}
with the events $N$ and the total energy $m$. For the sake of simplicity, the background from charmonium states is considered as the higher order correction and is set to be zero in a leading order analysis.

\begin{figure}[htb]
	\centering
 \includegraphics[scale=0.80]{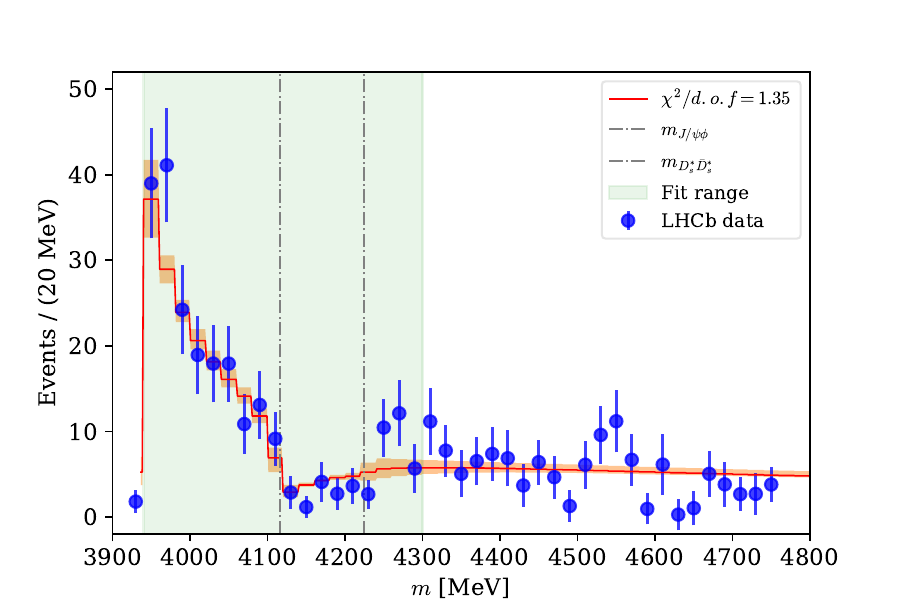}
	\caption{The best fit to the lineshape in $D_s\bar{D}_s$ invariant mass distribution including $\{D_s\bar{D}_s,\, J/\psi\phi,\,D_s^{\ast}\bar{D}_s^{\ast}\}$ coupled channel scattering in $J^{PC}=0^{++}$ sector. The orange bands represent the uncertainty (in $1\,\sigma$)  of the fitted lineshape.}\label{Fit}
\end{figure}

\begin{table}[htb]
\begin{tabular}{|c|c|c|}
\hline
$a_{11} \left[ \rm{fm}\right]$ & $a_{22} \left[ \rm{fm}\right]$ & $a_{23} \left[ \rm{fm}\right]$\\
\hline
$1.48\pm 0.17$ & $1.11\pm 0.65$ & $0.87\pm 0.10$\\
\hline
$P_1$ & $P_3$ & $a_{22}^{eff}\left[ \rm{fm}\right]$ \\
\hline
$-0.29\pm 0.02$ & $0.04\pm 0.03$ & $0.12^{+0.20}_{-0.10}+i0.78^{+0.20}_{-0.40}$\\
\hline
\hline
(-++) & (-++) & (-\,-+)\\
\hline
$3904.44^{+5.21}_{-14.68}$ & $4106.09^{+10.42}_{-58.26}-i 24.56^{-6.73}_{+29.14}$ & $4223.51^{+1.85}_{-4.29}-i 4.32^{-0.74}_{+14.49}$\\
\hline
\end{tabular}
\caption{The fit parameters and predicted poles corresponding to the lineshape in \ref{Fit}. The units of  $P_i$ and poles are  $\rm{MeV}^{-1/2}$ and $\rm{MeV}$, respectively. The sign +/- stands for the first/second Riemann sheet of the scattering channel.}\label{tab:table2}
\end{table}


The scattering amplitude is fitted in the Table. \ref{tab:table2} are expanded around $J/\psi \phi$ threshold, and different with the physical scattering lengths of $D_s\bar{D}_s$ and $D_s^{\ast}\bar{D}_s^{\ast}$, where the differences corresponding to the correction in loops are attributed to the subleading interaction in $\mathcal{O}\left(q^2 \right)$ order and do not affect the poles. Five parameters are adopted in the fit model, and a $\chi^2/d.o.f=1.35$ is achieved in the fit region, where the lineshape is well reproduced. Interestingly, although the fit
was only performed up to $4300\,\rm{MeV}$, a good description of the data is achieved in the entire energy interval up to
$4750\,\rm{MeV}$. 

It should be emphasized that, given the limited statistics of current experimental data, a purely unconstrained mathematical scan of the parameter space might yield alternative local minima with comparable $\chi^2/d.o.f$ values. However, in our analysis, the parameter optimization is strictly guided by underlying physical constraints. Specifically, the dynamically generated pole near the $D_{s}\bar{D}_{s}$ threshold is required to be located on the second Riemann sheet, which corresponds to an attractive interaction that is insufficient to form a bound state. Furthermore, the single-channel $J/\psi \phi$ interaction is also attractive but does not form a bound state. Under these stringent physical requirements, the obtained parameter set representing $\chi^2/d.o.f = 1.35$ is the unique optimal minimum, yielding the most physically consistent interpretation of the coupled-channel dynamics.

 There are three scattering lengths $\left\{a_{11},\,a_{22} ,\, a_{23} \right\}$ with positive values in Table. \ref{tab:table2}. $a_{11}$ indicates the $D_s \bar{D}_s$ interaction around $ J/\psi \phi $ threshold, and is attractive. This interaction is consistent with $a_{11}^{Ji}=\left(9.5^{+1.5}_{-1.0}\right)\times 10^{-3} \,\rm{MeV^{-1}}$ (or $1.87^{+0.30}_{-0.25}\,\rm{fm}$) with $\Lambda=1000\,\rm{MeV}$ in the Gaussian regulator of the loop \cite{Ji:2022vdj}, where $a_{11}^{Ji,\,-1}=\frac{2\pi}{\mu_1}\left( C_0^{-1} + \mu_1 \Lambda/\left( 2\pi\right)^{3/2}
\right)$ with $C_0=\frac{1}{2}\left( \mathcal{C}_{0a}+ \mathcal{C}_{1a}\right)$ in single channel scattering. This match on $D_s \bar{D}_s$ scattering length allows to have a molecular $D_s\bar{D}_s$ \cite{Chen:2022dad,Mutuk:2022ckn,Abreu:2023rye,Liu:2025wwx}, named $X(3960)$.  
It is important to emphasize that the underlying structural nature of the $X(3960)$ remains under active debate~\cite{Chen:2022dad,Mutuk:2022ckn,Abreu:2023rye,Liu:2025wwx}. In our unitarized scattering amplitude framework, the $X(3960)$ emerges as a $D_{s}\bar{D}_{s}$ virtual state (molecular state), which phenomenologically drives the near-threshold enhancement observed in the invariant mass spectrum. It is worth noting that if such an enhancement is parameterized using a Breit-Wigner formula, it may mimic the behavior of a genuine resonance. In fact, if a genuine bare resonance, such as the $X(3930)$, is situated extremely close to the $D_{s}\bar{D}_{s}$ threshold, both the intrinsic resonance pole and the continuum scattering amplitude will strongly interfere and contribute to the final spectrum. Therefore, our interpretation of the $X(3960)$ as a dynamically generated state is fully compatible with the complex threshold kinematics discussed in recent literature.
The $a_{22}$ characterizes the $J/\psi \phi$ scattering length, and its uncertainty is relatively larger than the other scattering lengths owing to the $2^{++}$ pole near $J/\psi \phi$ threshold is not included in the fit, which is hard to be improved according to current events.
Moreover, $D^{\ast}\bar{D}^{\ast}$ channel is not included in the coupled channel scattering directly, and its effect is absorbed into the uncertainty of $a_{22}$ in the form of
\begin{eqnarray}
    \delta a_{22}^{D^{\ast}\bar{D}^{\ast}}&=&\frac{\vert \langle X(3960) | D^{\ast}\bar{D}^{\ast}\rangle\vert ^2}{\vert \langle X(3960) | D_s \bar{D}_s\rangle \vert ^2}a_{11}=\frac{\vert g_{X(3960), D^{\ast}\bar{D}^{\ast}}\vert ^2}{\vert g_{X(3960), D_s \bar{D}_s}\vert ^2}a_{11} =0.30 \pm 0.03 \, \rm{fm}, 
\end{eqnarray}
where $g_{X(3960), D_{(s)}^{(\ast)} D_{(s)}^{(\ast)}}$ are the central valued couplings in Ref. \cite{Ji:2022vdj}. The error of $\delta a_{22}^{D^{\ast} D^{\ast}}$ is mild compared to the center value and switched off in the following sections, which means the error on $a_{22}$ from $J/\psi \phi$ channel is $\delta a_{22}^{J/\psi \phi}=\delta a_{22}-\delta a_{22}^{D^{\ast}\bar{D}^{\ast}}=0.36\,\rm{fm}$.



\subsection{On the OZI suppressed interactions and  poles}

The $a_{22}$ characterizes the OZI suppressed interaction and is determined to be $0.82\pm0.49\,\rm{fm}$ in K-matrix fit from LHCb collaboration \cite{LHCb:2022aki}. This type of interaction expresses the multi-gluon exchange potential and relates to recent studies on the fully heavy hadron spectrum. The di-$J/\psi$ scattering is an OZI suppressed process and the scattering length in $0^{++}$ or $2^{++}$ sector is extracted to be larger than $0.99\,\rm{fm}$ in a three-channel model using $\{J / \psi J / \psi, \psi(2 S) J / \psi, \psi(3770) J / \psi\}$ \cite{Dong:2020nwy}. A similar scattering length is imposed to be $1.64-2.25 \,\rm{fm}$. Besides the observations on fully tetraquark states, there are predictions on OZI suppressed scattering from lattice simulations, and the scattering lengths are $1.43 \pm 0.23 \, \rm{fm}$ in $N\phi$ scattering  \cite{Lyu:2022imf},
$1.57\pm 0.08 \,\rm{fm}$ in di-$\Omega_{ccc}$ scattering \cite{Lyu:2021qsh} and $-0.18 \pm 0.02 \,\rm{fm} $ in di-$\Omega_{bbb}$ scattering \cite{Mathur:2022ovu}. These scattering lengths are in the window of 1 to 2 $\rm{fm}$ apart from the one in di-$\Omega_{bbb}$ scattering and overlap over the $a_{22}$. Such numerical overlaps suggest that while multi-gluon exchange provides a smooth non-resonant background for OZI-suppressed interactions, a more granular analysis is required to distinguish their specific dynamical origins.

It is instructive to compare the extracted $a_{22}$ with the $N\phi$ scattering length, which is determined to be $1.43 \pm 0.23$ fm from lattice simulations \cite{Lyu:2022imf}. 
Although their magnitudes are comparable, one should be mindful that the sign conventions for the scattering length definition may vary between different frameworks. 
Despite the numerical similarity, the underlying dynamical mechanisms are distinct. 
Specifically, the elastic $J/\psi\phi$ scattering lacks triangle diagram contributions similar to those found in the $N\phi$ case. 
Instead, the interaction in the $J/\psi\phi$ channel is notably influenced by resonances near associated thresholds, such as the $X(6200)$ \cite{Dong:2020nwy,Liang:2021fzr,Dong:2021lkh,Nefediev:2021pww,Wang:2022jmb,Niu:2022jqp,Song:2024ykq,Liu:2025nze} near the $J/\psi J/\psi$ threshold and the $f_{2}(2010)$ near the $\phi\phi$ threshold. 
These states provide a unique source for the scattering length in the $J/\psi\phi$ sector, distinguishing its dynamics from the meson-exchange or triangle-diagram mechanisms typically found in $N\phi$ scattering \cite{Yan:2026srb}.


Comparing to the result of $J/\psi \phi$ scattering in lattice simulation,
this value is slightly larger than $a_0=0.242\pm 0.041\,\rm{fm}$ extracted in lattice simulation with $m_{\pi}=156\,\rm{MeV}$  \cite{Ozaki:2012ce}. However, the effective scattering length of $J/\psi \phi$ in the coupled channel scattering is evaluated to be $a_{22}^{eff}= 0.12^{+0.20}_{-0.10}+i0.78^{+0.20}_{-0.40} \, \rm{fm}$ and its real part is similar to $a_0$. Regarding $a_{23}/a_{11}$ is around $0.57$ with ignoring the mass difference in the channel thresholds, the coupled channel scattering plays a crucial role in generating an effective interaction near $J/\psi \phi $ threshold, which is consistent with the conclusion in Ref. \cite{Ozaki:2012ce}, where no pole is found in $J/\psi \phi$ scattering. It is natural to conclude from the OZI-suppressed interaction that coupled channel scattering (Fierz rearrangement) is essential. This interaction can also be expressed in the language of EFT, which states that the coupled channel effect promotes the OZI-suppressed interaction, which deserves future studies on the theoretical and phenomenological sides.

According to the fit parameters, three poles are predicted, with errors evaluated from the lower (upper) limits of $a_{ij}$ taken simultaneously.
The first pole is on the second RS of the $D_s\bar{D}_s$ channel and is identified as a virtual pole.
The second and third poles are located on the first RSs of $J/\psi \phi$ and $D_s^{\ast}\bar{D}_s^{\ast}$ thresholds, respectively,  where the coupled channel effect, from the higher channel $D_s^{\ast}\bar{D}_s^{\ast}$, enhances the attractive potential and generates a pole closing to the physical RS of $J/\psi \phi$. This pole performs as a resonance.
On the contrary, the inelastic transition between $J/\psi \phi$ and $D_s^{\ast} \bar{D}_s^{\ast}$ reduces the strength of the effective potential in the elastic $D_s^{\ast}\bar{D}_s^{\ast}$ scattering, and pushes the pole to be away from the threshold on the (-~-~-) RS, which is a general behavior among poles \cite{Yan:2018zdt,Peng:2023lfw,Wang:2025jcq}. However, the third pole is predicted on the (-~-~+) RS, corresponding to a stronger attractive potential than the one related to a pole on the (-~-~-) RS. Such a discrepancy can be understood from the error estimation of this type of EFT, which is in the order of $\mathcal{O}(\delta_0)$. When the error corresponds to enhancing the effective potential, the virtual pole may move to the physical sheet.

Apart from the S-wave scattering in $D_s\bar{D}_s$ invariant mass distribution,
 $Y(4230)$ plays a crucial role in generating the third pole, where the S-P coupled channel scattering $\{D_s^{\ast}\bar{D}_s^{\ast}, \, D_1(2420)\bar{D} \}$ can be referred to the $\bar{D}^{\ast}\Sigma_c-\bar{D}\Lambda_{c1}(2595)$ scattering \cite{Peng:2020gwk}, and is not covered in this work. Moreover,
the $Y(4230)$ provides a hint to understand the difference in the magnitudes of $P_1$ and $P_3$, which conflicts with the HQSS. The large uncertainty of $P_3$ suggests the behavior of lineshape around $D_s^{\ast}\bar{D}_s^{\ast}$ is not defined well enough in the current model, and will be improved with high statics in this region.


There is another choice to explain the attractive interaction in an OZI suppressed process, where a bare pole near $J/\psi \phi$ threshold promotes the interaction, which is the new $X_0(4140)$ resonance fitted in the form of Breit-Wigner from LHCb. This configuration naturally mixes with the dynamics of coupled channel scattering. To separate these two pictures of $X_0(4140)$, the studies on $J/\psi \phi$ scattering in different $J^{PC}$ are necessary, which relates to the enhancement ($X_1(4140)$) around $J/\psi \phi$ threshold in $J^{PC}=1^{++}$ sector.

\section{$X_{s\bar{s}}$ in $J^{PC}=1^{++}$ sector}


In addition to the coupled channel scattering in $J^{PC}=0^{++}$ sector,
$J/\psi \phi-D_s\bar{D}_s^{\ast}$ scattering, stemming from HQSS, may generate  poles in $J^{PC}=1^{++}$ sector. The scattering length in the $J/\psi \phi$ scattering is the same as in the $J^{PC}=0^{++}$ sector, where the spin-spin interaction is subleading.
In $D_s^{\ast}\bar{D}_s + h.c.$ scattering, the wave function is defined as 
\begin{eqnarray}
    X\left( 1^{++}\right)&=&  \frac{1}{\sqrt{2}}\left(\vert D_s^{\ast} \bar{D}_s\rangle + \vert D_s \bar{D}_s^{\ast}\rangle \right),
\end{eqnarray}
which generates the interaction $V_{D_s^{\ast}\bar{D}_s,\,D_s^{\ast}\bar{D}_s}=\frac{1}{2}V_{D_s \bar{D}_s,\,D_s \bar{D}_s}$, 
where the spin-spin interaction is relatively suppressed and also in subleading terms. This interaction is process dependent and couples to the $\vert 1_l \otimes 1_Q\rangle$ component.
In the inelastic transition,
the wave function of $D_s^{\ast }\bar{D}_s$ can be decomposed according to HQSS, 
\begin{eqnarray}
    \left(\begin{array}{l}
\left|D_{s}^{\ast} \bar{D}_{s}\right\rangle \\
\left|D_{s}\bar{D}_{s}^{\ast}\right\rangle  \\
\left|D_{s}^* \bar{D}_{s}^*\right\rangle
\end{array}\right)_{J=1}=\left(\begin{array}{ccc}
-\frac{1}{2} & \frac{1}{2} & \frac{1}{\sqrt{2}} \\
\frac{1}{2} & -\frac{1}{2} & \frac{1}{\sqrt{2}}\\
\frac{1}{\sqrt{2}} & \frac{1}{\sqrt{2}} & 0
\end{array}\right)\left(\begin{array}{l}
|0_l \otimes 1_Q\rangle \\
|1_l \otimes 0_Q\rangle \\
|1_l \otimes 1_Q\rangle 
\end{array}\right)_{J=1},
\end{eqnarray}
which yields a relative strength between $D_s\bar{D}_s^{\ast}$ and $D_s^{\ast}\bar{D}_s^{\ast}$ transiting into $J/\psi \phi$,
\begin{eqnarray}
    V_{J/\psi \phi, D_s \bar{D}_s^{\ast}}&=&\sqrt{2}V_{J/\psi \phi,\bar{D}_s^{\ast}\bar{D}_s^{\ast}}.
\end{eqnarray}

Concerning $D^{\ast}_{(s)}\bar{D}_{(s)}^{\ast}$ do not couple to $J/\psi \phi$ in $J^{PC}=1^{++}$ sector, a virtual pole $4106.74^{-4.02}_{+15.51}- i\, 8.26^{+2.56}_{-6.18}\,\rm{MeV}$ appears near $J/\psi \phi$ threshold. 
This narrow pole generates a sharp enhancement around the threshold in $J/\psi \phi$ invariant mass distribution, which is consistent with the observations on $X_1(4140)$ \cite{CDF:2009jgo,LHCb:2012wyi,CMS:2013jru,D0:2015nxw,LHCb:2021uow}. Besides this pole, there is a virtual pole ($4060.09^{+6.61}_{-3.99}\,\rm{MeV}$) strongly coupling to $D_s^{\ast} \bar{D}_s$ as a partner of molecular $\chi_{c1}(3872)$ aka $X(3872)$.

\begin{figure}[htb]
	\centering
 \includegraphics[scale=0.550]{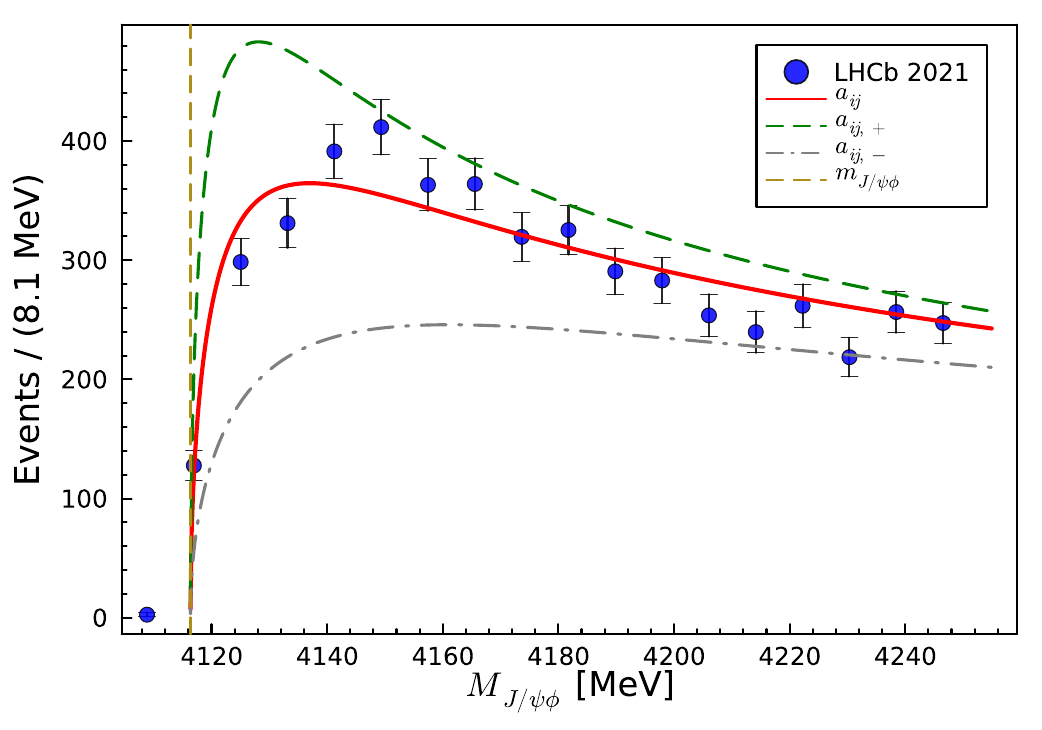}
	\caption{The prediction of lineshape of $X_1(4140)$ in $J/\psi \phi$ invariant mass distribution with an $a^{eff}$. The dash-dotted, solid and dashed  lines correspond to choices of the lower limits ($a_{ij,-}$), central value ($a_{ij}$) and upper limits ($a_{ij, +}$) of the parameters, which are taken to generate $a^{eff}$. The mass and width are parameterized to be $4118 \pm 11_{-36}^{+19}$ and $162 \pm 21_{-49}^{+24} \, \rm{MeV}$ in the Breit-Wigner profile, respectively.}\label{X1predicted}
\end{figure}
\begin{figure}[htb]
	\centering
 \includegraphics[scale=0.50]{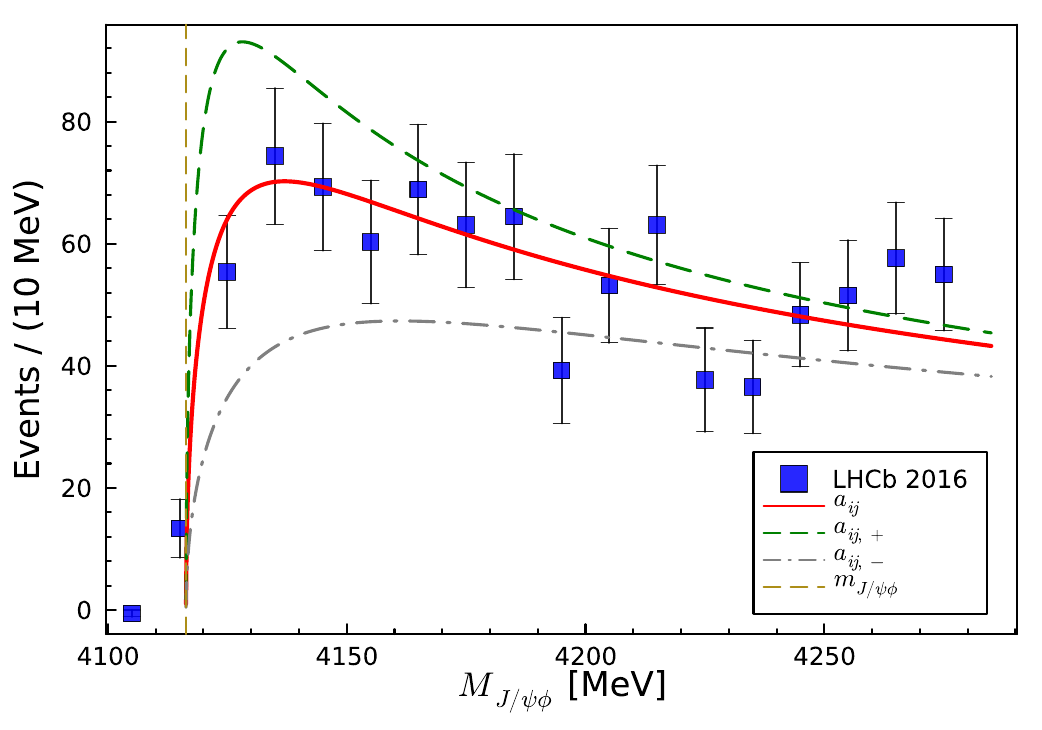}
	\caption{The prediction of lineshape of $X_1(4140)$ in $J/\psi \phi$ invariant mass distribution with an $a^{eff}$, where the $a^{eff}$ are the same with the ones in Fig. \ref{X1predicted}, and the efficiency corrected data are quoted from Ref. \cite{LHCb:2016nsl} with $P_{1^{++}}=0.50\, \rm{MeV^{-1/2}}$. 
 The mass and width of $X_1(4140)$ are fitted to be $4146.5 \pm 4.5_{-2.8}^{+4.6}$ and $ 83 \pm 21_{-14}^{+21} \, \rm{MeV}$, respectively.}\label{X1predicted2016}
\end{figure}
\begin{figure}[htb]
	\centering
 \includegraphics[scale=0.50]{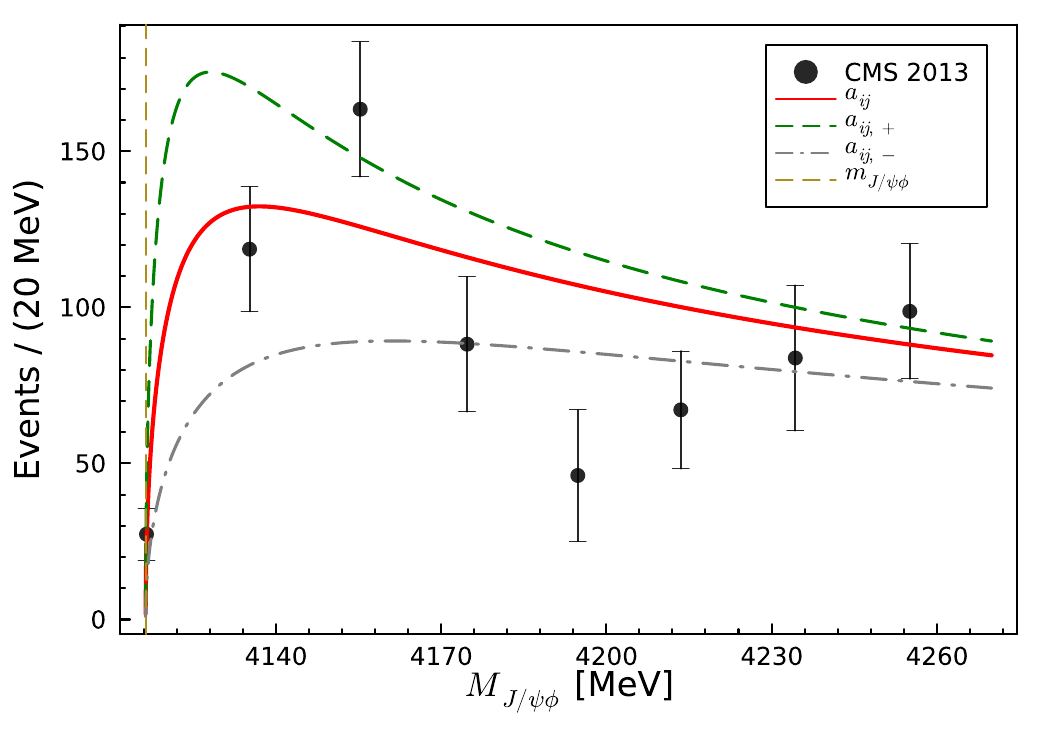}
	\caption{The prediction of lineshape of $X_1(4140)$ in $J/\psi \phi$ invariant mass distribution with an $a^{eff}$. The dash-dotted, solid and dashed lines correspond to choices of the lower limits ($a_{ij,-}$), central value ($a_{ij}$) and upper limits ($a_{ij, +}$) of the parameters, which are taken to generate $a^{eff}$ with $P_{1^{++}}=0.68\, \rm{MeV^{-1/2}}$. The mass and width are parameterized to be $4148.0 \pm 2.4$ and $28^{+15}_{-11} \, \rm{MeV}$ in the Breit-Wigner profile \cite{CMS:2013jru}, respectively.}\label{X14140PredictedCMS2013}
\end{figure}

In addition to the pole, the production of $X_1(4140)$ provides a hint to understand that state. Following the estimation on production in $J^{PC}=0^{++}$ sector, $P_{D_s^{\ast}\bar{D}_s}\gg P_{J/\psi \phi}$ holds in $J^{PC}=1^{++}$ as well. But $D_s^{\ast }\bar{D}_s$ threshold is lower than the one of $J/\psi \phi$ and $P_{D_s^{\ast}\bar{D}_s}$ is suppressed in $J/\psi \phi$ invariant mass distribution, which implies that $P_{J/\psi \phi}$ contributes to $J^{PC}=1^{++}$ spectrum dominantly. 
The amplitude of production of $B^-\to J/\psi \phi K^-$ in $J/\psi \phi$ invariant mass distribution is 
\begin{eqnarray}
    \mathcal{M}_b&=&P_{1^{++}} \frac{th_2}{8\pi} \frac{1}{1/a^{eff}-i k_2},
\end{eqnarray}
with the production rate of $J/\psi \phi$ $P_{1^{++}}$ and the effective scattering length $a^{eff}$ in coupled channel scattering,
which generates a distribution in $J/\psi \phi$ invariant mass
\begin{eqnarray}
\frac{d N}{d M_{J/\psi \phi}}&=& \vert \mathcal{M}_b\vert ^2 \frac{\vert \vec{k}_2\vert}{8\pi M_{J/\psi \phi}},
\end{eqnarray}
with the total energy $M_{J/\psi \phi}$. This distribution is performed with a fixed $P_{1^{++}}=1.16\, \rm{MeV^{-1/2}}$ in Fig. \ref{X1predicted}, where $P_{1^{++}}$ is chosen to match the tail of LHCb data points \footnote{Efficiency-corrected data were provided by Prof. Li-Ming Zhang.}. According to the choice on $P_{1^{++}}$, the predicted lineshapes cover the experimental observation well. Similar behaviors are shown in other observations in $J/\psi \phi $ invariant mass distributions from LHCb and CMS, presented in Figs. \ref{X1predicted2016} and \ref{X14140PredictedCMS2013}.

Comparing the Figs. \ref{X1predicted} and \ref{X1predicted2016}, the lines of theoretical predictions match the data from LHCb, where the narrow and broad $X_1(4140)$ are reported with respect to the rectangular and solid circled points, respectively. Interestingly, a set of dynamical parameters generating the curves of the LHCb observations does not meet the ambiguities of the $X_1(4140)$ widths in the Breit-Wigner formulae. To unveil the nature of $X_{1}(4140)$, this set of dynamical parameters is used to describe the observation from CMS \cite{CMS:2013jru}, where a narrower $X_1 (4140)$ is reported with $N=2480\pm 160$ events and a confidential level of 5 $\sigma$. The predicted lineshape is presented in Fig. \ref{X14140PredictedCMS2013}. The curves cover the data points as well. Such a coincidence is beyond the description from Breit-Wigner formulae, where the sharp change in $J/\psi \phi$ phase space is not included in the Briet-Wigner amplitude. Therefore, the enhancement from a near-threshold pole is a good candidate of the $X_1(4140)$, where the imaginary part of the pole is small and does not relate to the \textit{width} of the peak. The short distance between the real part of the pole and the threshold enhances the lineshape near the threshold. 

In the $1^{++}$ sector, the lower pole (located near the $D_{s}\bar{D}_{s}^{*}$ threshold) is further away from the $J/\psi \phi$ threshold than the higher pole. Therefore, the partner of $X(3872)$ is hard to detect in $J/\psi \phi$ invariant mass distribution. Alternatively, with $\phi$ decaying into $\rho \pi$ and 3 $\pi$, the lower pole can decay into  $J/\psi \pi\pi\pi$ and its signal may be smeared by the broad $\rho$ meson. Consequently, detecting the molecular $D_s^{\ast}\bar{D}_s$ state is challenging.




\section{$X_1(4685)$ in $B\to J/\psi \phi K$}

The parameter $a_{22}$ represents the multi-gluon exchange interaction between $J/\psi$ and $\phi$, and
 the dynamics in low energy $\psi(2S)\phi$ scattering coincide with those in $J/\psi\phi$ scattering. 
$\left\{ J/\psi \phi,\, \psi(2S)\phi\right\}$ coupled channel scattering involves a large mass difference between thresholds that is a new scale $\Lambda_{CC}=\sqrt{2\mu \left(m_{\psi(2S)}-m_{J/\psi}\right)}= 951.50\,
\rm{MeV} \simeq \Lambda_0$ \cite{Valderrama:2012jv,Yan:2021nio}, which implies the coupled channel effect is suppressed by $\mathcal{O}\left(\vec{k}_2^2/\Lambda^2_{CC} \right)$. The EFT should be expanded to the subleading term, $\mathcal{O}(\vec{k}_2^2)$ \cite{Albaladejo:2015lob}, which can not be extracted from the fit in Fig. \ref{Fit}. On the other hand, the interaction in the inelastic scattering is phenomenologically approached by
\begin{eqnarray}
    \mathcal{L}_{J/\psi \phi_{i},\, \psi(2S)\phi_{f}}&=& D_0\frac{E_{\phi_i}}{m_{\phi_f}}\vec{\epsilon}_{J/\psi}\cdot \vec{\epsilon}_{\psi(2S)} \vec{\epsilon}_{\phi_i} \cdot \vec{\epsilon}_{\phi_f},
\end{eqnarray}
where the relativistic $\phi$ meson is corrected by the energy dependent term $E_{\phi}$ \cite{Valderrama:2018knt}.

In the coupled channel scattering, the virtual poles may generate in $\psi(2S)\phi$ scattering in $J^{PC}=0^{++}\, ,1^{++}$ and $2^{++}$ sectors and transit to $J/\psi \phi$, 
\begin{eqnarray}
    0^{++}:&& \, J/\psi \phi [^1S_0],\,\psi(2S)\phi[^1S_0],\\
    1^{++}: &&\, J/\psi \phi [^3S_1],\,\psi(2S)\phi[^3S_1],\\
    2^{++}:&& \, J/\psi \phi [^5S_2],\,\psi(2S)\phi[^5S_2],
\end{eqnarray}
where the amplitude is unitarized to a $2\times 2$ t-matrix and contributes to the lineshape in $J/\psi \phi$ invariant mass distribution with the amplitude of transition
\begin{eqnarray}
  \mathcal{M}_c  &=&P^{J/\psi \phi}t_{11} + P^{\psi(2S)\phi}t_{21}
\end{eqnarray}
with subscripts 1 and 2 labeling $J/\psi \phi$ and $\psi(2S)\phi$, respectively. If $P^{\psi(2S)\phi} >P^{J/\psi \phi}$, the lineshape around $\psi(2S)\phi$ threshold relates to the modulus of $t_{21}$ and performs as a peak.

\begin{figure}[hbt]
	\centering
 \includegraphics[scale=0.50]{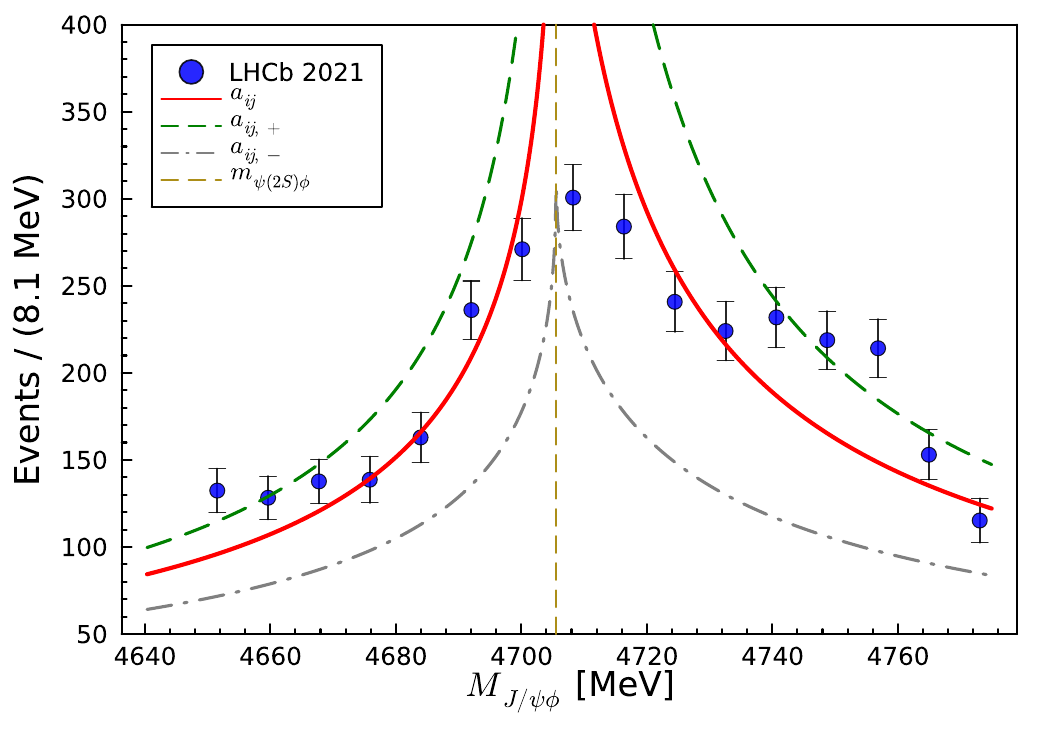}
	\caption{The prediction of lineshape of $X_1(4685)$ in $J/\psi \phi$ invariant mass distribution with an $a^{eff}$, is an analogy of the one in $X_1(4140)$ production. The dash-dotted, solid and dashed lines correspond to choices of the lower limits ($a_{ij,-}$), central value ($a_{ij}$) and upper limits ($a_{ij, +}$) of the parameters, which are taken to generate $a^{eff}$ with $P_{1^{++}}^{\psi(2S)\phi}=0.48\, \rm{MeV^{-1/2}}$. }\label{X14685PredictedLHCb2021}
\end{figure}

Concerning the enhancement near $J/\psi \phi$ threshold is mainly dominated from $1^{++}$ sector, the production rate of $J/\psi \phi$ in $1^{++}$ sector is larger than the ones in $0^{++}$ and $2^{++}$ sectors, which also happens in $\psi(2S)\phi$ production. Consequently, the $1^{++}$ virtual pole, $4690.85^{-5.73}_{+5.08}-i7.00^{+13.50}_{-7.42}\,\rm{MeV}$, generates a peak at $ \psi(2S) \phi$ threshold and is the candidate of the newly reported $X_1(4685)$, the fourth peak in $J/\psi \phi$ invariant mass distribution \cite{LHCb:2021uow}. This process is similar to the one in $J/\psi J/\psi- J/\psi \psi(2S)$ \cite{LHCb:2020bwg,ATLAS:2023bft,CMS:2023owd}. However, a series of charm meson pairs may couple to $J/\psi \phi-\psi(2S)\phi$ scattering, where the opened charm meson channels are saturated to the width of the peak that is not included in the model. Therefore, instead of evaluating  $X_1(4685)$ width directly, the lower limit of the half maximum width is evaluated as
\begin{eqnarray}
   \Gamma_{cusp}=60.90^{+62.54}_{-24.24} \, \rm{MeV}, 
\end{eqnarray}
according to the formulae,
\begin{eqnarray}
  \Gamma_{cusp}&=&  \frac{1}{\mu}\left(\frac{4}{\left|a_0\right|^2}-\sum_x x \sqrt{\frac{3}{\left|a_0\right|^2}+x^2}\right), \label{eq:GamCusp}
\end{eqnarray}
with the $\psi(2S) \phi$ reduced mass $\mu$ and the effective scattering length $a_0=1.03^{-0.39}_{+0.38}+i \, 0.40^{+0.01}_{-0.01}\,\rm{fm}$,
where the sum runs over $x=\operatorname{Im}\left(1 / a_0\right)$ and $\operatorname{Re}\left(1 / a_0\right)$ \cite{Dong:2020hxe}. 
Surprisingly, the predicted $\Gamma_{cusp}$ closes to the parameters, $\left(M,\,\Gamma_{BW}\right)=\left(4684 \pm 7_{-16}^{+13}, \, 126 \pm 15_{-41}^{+37} \right)\,\rm{MeV}$, extracted from the Breit-Wigner approximation \cite{LHCb:2021uow}. The predictions on the lineshapes are displayed in Fig. \ref{X14685PredictedLHCb2021} with a fixed production rate.
This coincidence supports that $X_1(4685)$ is a $\psi(2S)\phi$ molecule in addition to the threshold effect \cite{Nakamura:2021bvs,Luo:2022xjx}.

Besides $X_1(4685)$, there is a $X_0(4700)$ with quantum numbers $J^{PC}=0^{++}$, whose mass and width are very similar to the ones of $X_1(4685)$ in the Breit-Wigner approximation. This state can also couple to $\psi(2S)\phi$ in S-wave. However, the production rate of $J/\psi \phi$ in $J^{PC}=0^{++}$ needs to be larger, implying that this state is not visible in $J/\psi \phi$ invariant mass distribution. The $X_0(4700)$ may correspond to a different configuration.


\section{summary}

In summary, we have performed the first theoretical study of the near-threshold lineshape in the $D_s \bar{D}_s$ invariant mass distribution using an OZI-suppressed $\{D_s \bar{D}_s,\, J/\psi \phi, \, D_s^{\ast} \bar{D}_s^{\ast} \}$ coupled-channel scattering framework, where the effective range expansion is applied to the scattering amplitude. 

By fitting the experimental data with five parameters, we achieve a best fit with $\chi^2/d.o.f=1.35$. As discussed in detail in the main text (see Sec.~II~B), this optimal fit is uniquely determined by imposing rigorous physical constraints on the parameter space---such as requiring the near-threshold pole to lie on the second Riemann sheet in single channel scattering---which effectively excludes other mathematically possible but unphysical local minima. 

Based on this fully constrained amplitude, the effective scattering length of the $J/\psi\phi$ channel is extracted as $a^{eff}_{J/\psi\phi} = 0.12^{+0.20}_{-0.10}+i0.78^{+0.20}_{-0.40}\,\rm{fm}$. Furthermore, three dynamically generated poles in the $J^{PC}=0^{++}$ sector are found at $3904.44^{+5.21}_{-14.68}\,\rm{MeV}$, $4106.01^{+10.07}_{-58.18}-i 24.56^{-7.33}_{+29.14}\,\rm{MeV}$, and $4223.51^{+1.85}_{-4.29}-i 4.32^{-0.74}_{+14.49}\,\rm{MeV}$. 

The lowest pole naturally corresponds to the $X(3960)$ as a $D_{s}\bar{D}_{s}$ virtual state, and the second pole serves as a candidate for the molecular $X_{0}(4140)$. These dynamically generated states are fundamentally different from conventional bare resonances parameterized by simple Breit-Wigner formulas~\cite{LHCb:2022dvn}.

With the spin-spin interaction attributed to being subleading order, a virtual state is predicted below $J/\psi \phi$ threshold in heavy quark spin symmetry, which shows good performance to describe the observation near $J/\psi \phi$ threshold in $B^+\to J/\psi \phi K^+$ from LHCb and CMS. This coincidence means the $X_1(4140)$ with $J^{PC}=1^{++}$ could come from a virtual pole, and there is no more ambiguity on the widths of $X_1(4140)$. 
It is worth noting that our current framework is truncated at leading order to avoid over-parameterization given the present experimental statistics. 
The subleading $\mathcal{O}(q^2)$ corrections, including the spin-dependent interactions and effective range contributions, are implicitly absorbed into the theoretical uncertainties. 
With more precise measurements and larger data samples expected soon from the LHCb collaboration, these higher-order partial-wave contributions and coupled-channel dynamics can be explicitly isolated, which will provide a more rigorous test of the heavy quark spin symmetry and the exact locations of these near-threshold poles.


In addition to the spectrum of $X_{0}(4140)$, $X_1(4140)$ and $X_1(4685)$, the OZI-suppressed interaction can be promoted in the coupled channel scattering as a consequence of Fierz rearrangement. The updated observation in $D_s \bar{D}_s$ invariant mass distribution is valuable to gain a deep insight into the nonperturbative strong interaction.

\section*{ACKNOWLEDGEMENT}

M.J. Yan would like to thank Feng-Kun Guo, Jia-Jun Wu, Bing-Song Zou, Hong-Rong Qi, and Zhen-Hua Zhang for the valuable discussions. 
Special thanks are due to Prof. Li-Ming Zhang for providing the efficiency-corrected data from the LHCb collaboration. 
This research is supported by the National Natural Science Foundation of China under Grant No. 12305096, and the Fundamental Research Funds for the Central Universities under Grant No. SWU-KQ25016.

\bibliography{refs.bib}

\clearpage
\section*{Appendix I: ERE and effective potential}
According to inverse amplitude method (IAM) \cite{Dobado:1996ps,Nieves:1999bx}, a relation between contact range interactions and scattering lenghts in $J^{PC}=0^{++}$ sector is presented by a LO analysis,
\begin{eqnarray}
  T^{-1}=  V^{-1}-G &=& \left(\begin{array}{ccc}
V_{11} & V_{12} & V_{13} \\
V_{12} & V_{22} & V_{23}\\
V_{13} & V_{23} & V_{33}
\end{array}\right)^{-1}- \left(\begin{array}{ccc}
G_{11} & 0 & 0 \\
0 & G_{22} & 0\\
0 & 0 & G_{33}
\end{array}\right)
\end{eqnarray}
with subscripts labeling the channels and $V_{13}=0$ according to the power counting rule in this note, whose elements are
\begin{eqnarray}
    (T^{-1})_{11}&=& \frac{V_{22} V_{33}-V_{23}^2}{det}-G_{11}, \\
     (T^{-1})_{22}&=& \frac{V_{11}V_{33}}{det}-G_{22}, \\ 
         (T^{-1})_{33}&=& \frac{V_{11}V_{22}-V_{12}^2}{det}-G_{33}, \\ 
             (T^{-1})_{12}&=& \frac{V_{12}V_{33}}{det}, \\ 
             (T^{-1})_{23}&=& \frac{V_{11}V_{23}}{det}, 
\end{eqnarray}
with $det=-V_{33} V_{12}^2-V_{11} V_{23}^2+V_{11} V_{22} V_{33}$,
where $V_{im}V_{mj}$ in the numerators corresponding to loop correction is attributed to be the subleading term. According to $V_{11}=V_{33}$, the matrix turn to be
\begin{eqnarray}
  (T^{-1})_{11}&=& \frac{1}{V_{11}}-G_{11}, \\
     (T^{-1})_{22}&=& \frac{1}{V_{22}}-G_{22}, \\ 
         (T^{-1})_{33}&=& \frac{1}{V_{33}}-G_{33}, \\ 
             (T^{-1})_{12}&=& \frac{V_{12}}{V_{12}^2+V_{23}^2- V_{11}V_{22}}\sim -\frac{V_{12}}{V_{11}V_{22}}, \\ 
             (T^{-1})_{23}&=& \frac{V_{23}}{V_{12}^2+V_{23}^2- V_{11}V_{22}}\sim -\frac{V_{23}}{V_{11}V_{22}}.  
\end{eqnarray}
These matrix elements match the ones in $(1/a)_{ij}$. Inversely, the leading order $V_{ij}$ can be expressed by the $a_{ij}$. Note that the off-diagonal elements $\left(T^{-1} \right)_{12}$ and $\left(T^{-1} \right)_{23}$ involve a denominator of V-matrix, which relates to a coupled channel effect. Regarding the ERE is advocated around $J/\psi \phi$ threshold, $V_{11}V_{22}$ dominates the strength according to $V_{12}^2$ and $V_{23}^2$ are in subleading terms. 

\section*{Appendix II: triangle diagram in $B\to D_s\bar{D}_s K$}
\begin{figure}[htb]
	\centering
 \includegraphics[scale=0.50]{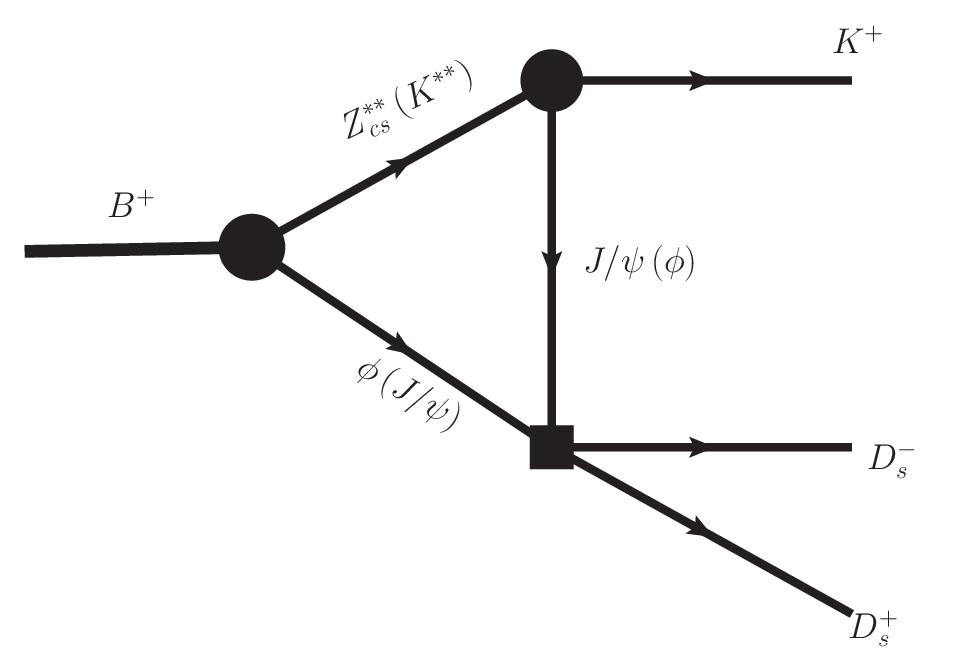}
	\caption{Two types of transitions in triangle diagrams driven by propagated $Z_{cs}^{**}\,\left( K^{\ast \ast}\right)$.}\label{TS}
\end{figure}

Besides the threshold effect from the two-body scattering, there are observable threshold effects from rescattering in triangle diagrams, named triangle singularity (TS),  \cite{Guo:2019twa}, and the processes are represented by the Feynman diagrams in Fig. \ref{TS}.
In the diagram driven by $Z_{cs}^{**}$ and $ K^{**}$, the TSs are fulfilled with mass windows, 
\begin{eqnarray}
   m_{Z_{cs}^{**}} &\in&  \left[ 4227.57\, , 4259.88\right] \, \rm{MeV},
   \\
   m_{K^{**}} &\in &  \left[ 1982.13 \, , 2182.44\right] \,\rm{MeV}.
\end{eqnarray}
There are $Z_{cs}(4220)$, $K_0^{\ast}(1950)$, $K_2^{\ast}(1980)$  and $K_4(2045)$ \cite{Workman:2022ynf} in these mass windows, whose widths are in the range of 100 to 350 $\rm{MeV}$.
Such broad states smear the TSs dramatically, similar to the smearing from meson widths in Refs. \cite{Liu:2013vfa,Liu:2020orv,Yan:2022eiy,Yan:2025bez}. Regarding $Z_{cs}(4220)$ and $K^{**}$ widths are much larger than the ones of $D^{\ast}$ and $D_1(2420)$, the TSs from Fig. \ref{TS} are heavily suppressed and not included, serving as a smooth background term in $D_s\bar{D}_s$ invariant mass distribution.

\section*{Appendix III: Hidden strange $X_2$ in $J/\psi \phi$ invariant mass distribution}
Concerning the observation of $X_1(4140)$ is the only one confirmed state near $J/\psi \phi$ threshold in $B\to J/\psi \phi K$ and the spin interaction in $J/\psi \phi$ scattering is subleading, the production rates of $J/\psi \phi$ may obey $P_{1^{++}}\gg P_{0^{++}}$ and $P_{1^{++}}\gg P_{2^{++}}$.
In $\{J/\psi \phi,\, D_s^{\ast}\bar{D}_s^{\ast} \}$ coupled channel scattering, the production rates of $J/\psi \phi$ are set to be $P_{0^{++}}= P_{2^{++}}=0$. The poles near $D_s^{\ast}\bar{D}^{\ast}_s$ threshold may generate lineshape with a nonzero production rate of the charm meson pair $P_{D_s^{\ast}\bar{D}_s^{\ast}}$.

The $0^{++}$ pole near $D_s^{\ast}\bar{D}_s^{\ast}$ threshold is listed in Table. \ref{tab:table2} and the one 
in $2^{++}$ sector in $\{J/\psi \phi,\, D_s^{\ast}\bar{D}_s^{\ast} \}$ coupled channel scattering has the decomposition of the wave functions,
\begin{eqnarray}
    \vert D_s^{\ast} \bar{D}_s^{\ast}\rangle_{J=2} = \vert 1_l \otimes 1_Q\rangle_{J=2} , && \,\,\,\,\vert J/\psi \phi \rangle_{J=2}= \vert 1_l \otimes 1_Q\rangle_{J=2}.
\end{eqnarray}
According to the decomposition of spin wave function of $D_s^{\ast}\bar{D}_s^{\ast}$, $\vert 1,0;1,0\rangle =\sqrt{\frac{2}{3}}\vert 2,0\rangle -\sqrt{\frac{1}{3}}\vert 0,0\rangle$, the relative strength of $J/\psi \phi -D_s^{\ast}\bar{D}_s^{\ast}$ between $J^{PC}=0^{++}$ and $2^{++}$ is 
\begin{eqnarray}
    V_{D_s^{\ast}\bar{D}_s^{\ast},\, J/\psi \phi}\left( 2^{++}\right)&=& \frac{\mathcal{N}_0}{\sqrt{2}}V_{D_s^{\ast}\bar{D}_s^{\ast},\, J/\psi \phi}\left( 0^{++}\right).
\end{eqnarray}
As a result,
there are virtual poles around $J/\psi \phi$ and $D_s^{\ast}D_s^{\ast}$ thresholds, respectively, the latter one of which contributes to the spectrum in $J/\psi \phi$ invariant mass distribution with the pole $4217.82^{+1.07}_{-1.46}+ i 3.11^{-0.49}_{+0.41} \,\rm{MeV}$, where $a_{12}^{J/\psi \phi,\, D_s^{\ast}\bar{D}_s^{\ast}}=\frac{1}{\sqrt{2}} a_{23}$ is adapted concerning HQSS.

 Assuming the production rates of $D_s^{\ast}\bar{D}_s^{\ast}$ in $0^{++}$ and $2^{++}$ sectors share a strength, the cross section in $2^{++}$ sector is larger than the one in $0^{++}$ sector in S-wave transition, which is proportional to $\vert V_{D_s^{\ast}\bar{D}_s^{\ast}, J/\psi \phi}\vert^2$.
Regarding the difference in thresholds of $D_s^{\ast}\bar{D}_s^{\ast}$ is $108 \,\rm{MeV}$, the D-wave transition is not negligible compared with the S-wave one in $J/\psi \phi$ invariant mass distribution. In addition, the three-momentum dependent D-wave transition causes a dip at $D_s^{\ast}\bar{D}_s^{\ast}$ threshold in the form of $t=c_1 k_2^2 + c_2 k_3^2$, which matches the new observations from LHCb \cite{LHCb:2021uow, LHCb:2023hxg}. The dip around $D_s^{\ast}\bar{D}_s^{\ast}$ threshold is linearly dependent on the invariant mass, which could be approached by 
\begin{eqnarray}
    \frac{dN_{J/\psi \phi}}{dm} &\propto&\vert k_3^2\vert ^2 k_2/m=\left[2\mu_3\left(m-2m_{D_s^{\ast}}\right) \right]^2 k_2/m,
\end{eqnarray}
with an approximated constant $k_2$ around $D_s^{\ast}\bar{D}_s^{\ast}$ threshold,
where the subscripts 2 and 3 refer to the channels $J/\psi \phi$ and $D_s^{\ast}\bar{D}_s^{\ast}$, respectively, $k_i$ stand for the momentum in the center of mass frame, and $\mu_i$ are the reduced masses.
 This is a further hint on the D-wave transition. Therefore, the dip is interpreted to be a $2^{++}$ molecular state in $J/\psi \phi-D_s^{\ast} \bar{D}_s^{\ast}$ scattering.


\end{document}